# Unified Modeling Language for Describing Business Value Chain Activities

Ashish Seth
Research Scholar,
Punjabi University, Patiala, India

Himanshu Agarwal
Associate Professor,
Punjabi University, Patiala India

Ashim Raj Singla
Associate Professor,
IIFT, New Delhi, India

## ABSTRACT

With the market competition aggravating, it becomes necessary for market players to adopt a business model which can adopt dynamic business changes. Any enterprise has the possibility to win in the competition only when it forms the strategic alliance with the upstream and downstream enterprise. This paper articulates a way of using unified modelling language (UML) to develop business value chain activities for any enterprise to develop dynamic, adhoc and agile business model. The results show that the UML is useful in the development of information systems and is independent of any programming language.

## 1. INTRODUCTION

It is really difficult to integrate the systems that are designed to work independently for attaining particular functionality done, for example SCM, CRM, and ERP etc these systems are not designed to work together. More generally, the integration between ERP and CRM can be defined as a medium to collaborate front-end and back office operation of an organization that applied them. They enable customers and business partner to be included into value chain inward to and outward from the organization, and encourage the collaboration between companies. It is also identified by Chong Kwong Chen (May2011) that in this era of competitive business world SCM, CRM and ERP has become the most influential enterprise systems in term of improving competitive advantages of an organization. But it is also found that these systems running in companies are continue to exist in isolation and become less relevant in today business context due to lack of integrated information achievable through each of the system respectively. The four major reasons that Chong K.C suggested are existing systems lack of functional interoperability, customer and supplier insufficiently collaborated with existing ERP, no standards or open standard methodology for information exchange and lack of interfacing mechanism or tool to cater system change.

Till date, many experts and researchers have put forward software models, but most of them are based on the traditional structured method, this paper uses UML and object-oriented analysis method to analyze business modeling value chain activities, the model can enhance the exchange among the experts, software designers and users, making system develop smoothly. According to Chang and Makatsoris (2001), "It is very hard for ERP systems to perform real time simulation of adjustments due to constraints in the manufacturing, finance, distribution, warehouse or other operations because the system is mainly concerning on the business transaction processing". Meanwhile, Robinson and Wilson (2001) stated that customer services must be directly connected to ERP systems in order to react rapidly to customer orders. It is important to generate a platform for direct communication between actual demand to production and planning (Robinson & Wilson, 2001).

Considering these researches, it is very obvious that any isolated system be it ERP, SCM, CRM, though very efficient in its own is not good enough to improve the competitive advantages due to lack of collaboration among the company, customers and suppliers in the business context. These systems need to be integrated with each other in order to cater with continuous changes in business over a period of time.

According to Arshah, Desa, and Hussin (2008), stakeholders do not have the knowledge of accessing to the internal workings of the system due to lack of support from experienced colleagues. Hence, many of the legacy systems were developed by using proprietary technologies and most of them are lack of standards like XML (Kuchibhotla, Dunn & Brown, 2009; Cruz & Rajendran, 2003).

The main objective of this work is to design and simulate a service based system that enables interoperability among ERP, CRM and SCM for any SME. In order to understand the overall integration we first discuss important concept that are used to represent the overall business values chain integrated architecture i.e. Unified Modeling Language (UML), Model Driven Architecture (MDA), Uses cases (UC) and Service Delivery Life Cycle (SDLC).

### 1.1 Unified Modeling Language (UML)

The Unified Modeling Language (UML) is a family of graphical notations, backed by a single meta-model that help in describing the designing software systems, particularly software systems built using the object oriented(OO)style. Graphical modelling languages have been around in the software industry for a long time. The fundamental driver behind them is that programming languages are not a high enough level of abstractions to facilitate discussions about designs. UML is a relatively open standard, controlled by the Object Management Group (OMG), an open consortium of companies. The OMG was formed to build standards that supported interoperability, specifically the interoperability of object oriented systems. The elements of the UML map pretty directly to elements in the software systems. UML has been identified as a way of providing a solution to modelling bottle neck [4].

## 1.2 Model Driven Architecture (MDA)

Model Driven Architecture (MDA) is the standard approach to using the UML as programming language; the standard is controlled by the OMG. By producing a modelling environment that conforms to the MDA, vendors can create models that can also work with other MDA complaints environments. MDA divides development work into two main areas- Platform Independent Model (PID) and Platform Specific Model (PSM). The PIM is a UML model that is independent of any particular technology, whereas PSM is a model of a system targeted to a specific execution environment. There are certain tools that can turn PIM into PSM. In this paper we have focused on to the design that is independent of any technology and presented the PIM model for integration of business values chain activities of any SME. Later this can be run on J2EE or .NET by using some vendor tools to create PSMs, tools would generate code for these platforms.

## 1.3 Use Case (UC)

Use case modeling is very popular within the software engineering community and service requirements can be effectively analyzed through use case modeling. Use case modeling makes the user understand how the system works through the relationships between actors and use cases. Use case modeling is user based and a function oriented analysis method. It is quite effective as the requirements analysis method.

Use-Case Analysis is a process of identifying the conceptual items and properties necessary for a solution to be both correct and proper. The use-case model describes what the system does for each type of user. The actors are entities that interact with the system and the use cases are complete functionalities as perceived by an actor. The use-case model provides essential input for analysis, design and testing. Use cases offer a systematic and intuitive way to capture the functional requirements with particular focus on the value added to each individual user or to each external system. Use cases play a key role in driving the rest of the development work and that is the important reason for their acceptance in most approaches to modern software engineering.

Example

The following diagram (see fig 1) shows a simple-first level use case diagram for simple buying and selling scenario where the Buyer and Sellers are the major actors of the system.

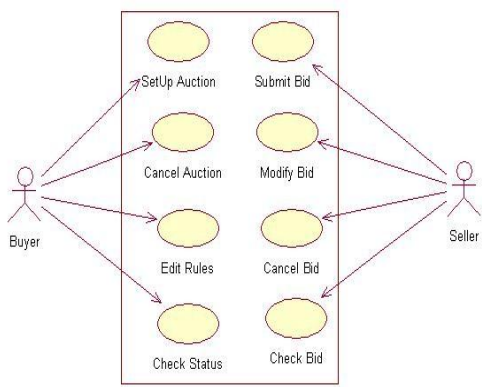

**Fig. 1 Use case for Buying-Selling Scenario**

## 2. Service Delivery Life Cycle (SDLC)

SDLC in context of SOA starts with Service oriented analysis followed by service oriented design, service development, service testing and finally service deployment. Throughout this process Service Administration is necessary to monitor the designed service and their orchestration for adhoc functionality (see figure 2). The task performed at each phase is as follows [14]:

- Service-oriented analysis, moreover, determines potential scope of SOA within the organization, Service are identified and mapped out from traditional legacy system to model as smart services

- In Service-oriented design phase, standards and protocols are designed conforming service level agreement (SLA), along with business processes.

- Service Development phase is actual construction phase where services identified in design phases are coded using suitable language

- Service Testing phase is required to undergo rigorous testing of services prior to deployment

- Service Deployment needs to configure distributed components, service interfaces, and any associated middleware products onto the production servers.

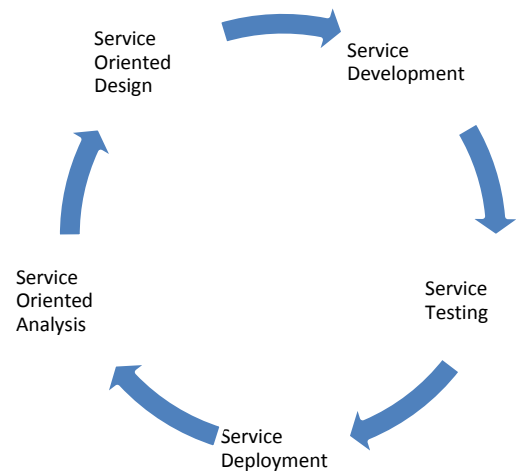

**Fig. 2 SOA Delivery Life Cycle**

Services Administration is needed from service development phase onwards to keep monitor the designed services and their orchestration for adhoc functionality

## 3. Customer Relationship Management (CRM)

In today's global trading environment, a flexible and robust customer data model is needed to capture customer's information to keep track of their customer's behaviour, to interact with them and prospect potential customers in order to forecast what their customer's trend will be in future. Not just commercial companies need these functions, but any institution that interacts with other institutions, such as universities, clubs, or social associations, though universities, social associations and clubs for example, do not really trade, they keep track of students information, need to attract them in many ways

There has been much work done in domain-specific areas, such as analysis patterns for Accounting [8], Reservations [9], and Course Management for educational settings [10], but none of them capture a generic model that can be applied to the entire trading community. The following section describes some aspects of recording information about customers for an organization in a trading community that sells item to its customers, item may be any product or any kind of service. A trading community is defined as a group of entities taking part in some type of commerce. It includes persons and organizations. Besides seller and buyer, entities in a trading community can be Partner, Contactractor, Distributor, Dealer, Agent, Influencer, etc. (see figure 3 below)

Companies or organizations need to interact with these entities to operate their business. Customer relationship has a broader context than classical customers, not only it represents the customer model; it also represents multiple organizations and multiple relationships that exist in a complex matrix-like environment [1].

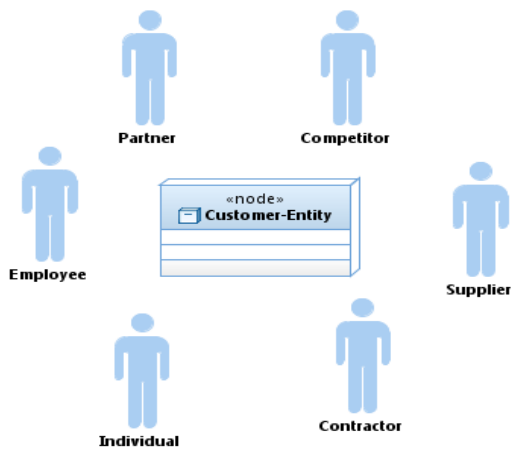

**Fig. 3 Customer entities in trading community**

A more generic customer model must reflect their prospective customer and their relation ship with them; these relationships may be dynamic and can change at any time. Use case diagram representing the model are depicted in [figure 4]. Actor can be any individual or a company, Relationship links to two entities to indicate the nature of relationship between them. Examples of such relationship are: supplier to distributor for, client of/contractor to, report to/ manager of, customer of seller to etc.

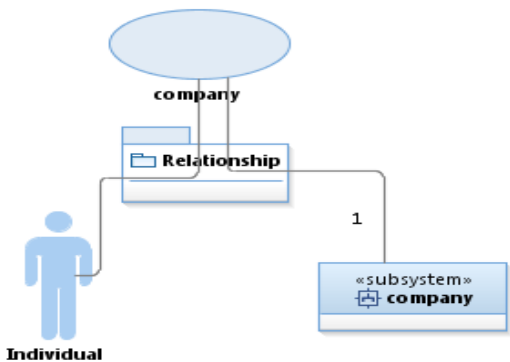

**Fig. 4 Use Case Diagram for Customer Relationship**

Use case diagrams address the business processes that the system will implement. The Use Case is further used to identify classes to be represenetd in a UML and helps in establishing a relationship among classes in class diagram. The following class diagram [figure 5] depicts the relationship between the two entities (buyer and seller). The company relationship may be either business to business (B2B) or business to customer (B2C)

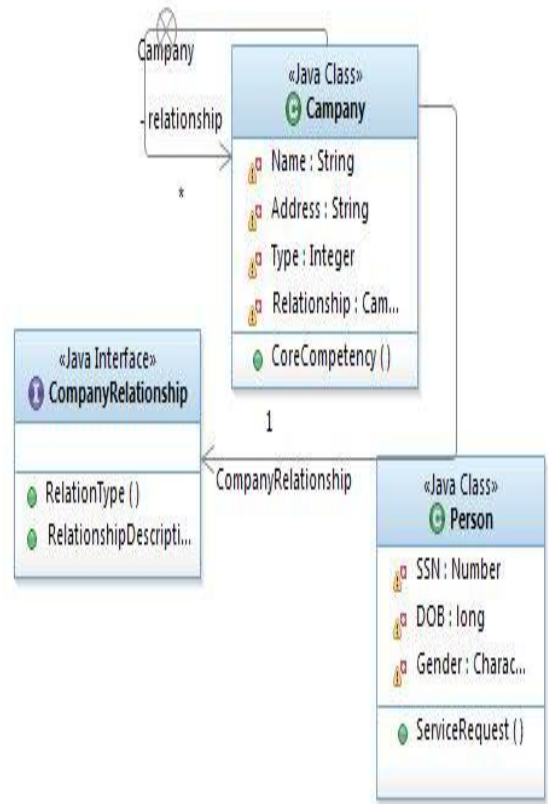

**Fig. 5 Class Diagram for Relationship among Business Entities**

After coming up with set of use cases, by examining the problem statement we come up with an analysis level class diagram which is as shown below[figure 6]. The classes are identified by examining the nouns in the statement. **Company** represents any business entities that can either an individual or an organization doing business. *relationship* links to itself indicating the entities involved between a business deal. **Location** is essentially the physical location of all such business entities. **CompanyRelationShip** links two business entities to indicate the nature of relationship between them, regardless of their type. Person is an individual entity who is involved in any sort of business and takes care of personal details. **CustomerInfo** uniquely identifies links through **CompanyRelationship**, *communicationPoint* is an identifier for a point of contact to an any sort of customer. **CustomerPattern** take care of keeping and identifying the customer behavior and keeping information about changing customer trends over time, it is linked though *CustPattren* through **CompanyRelation Ship**

19

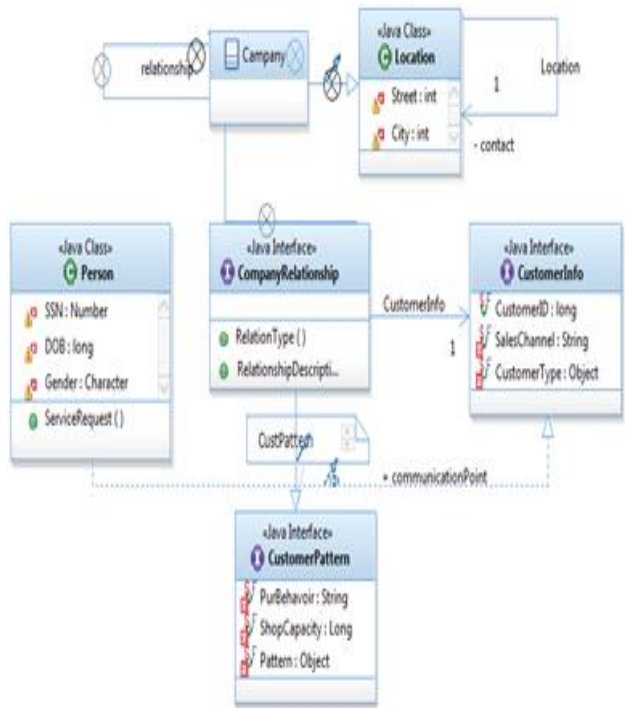

**Fig. 6 Class Diagram for Customer Relationship**

## 4. Supply Chain Management (SCM)

Supply chain management (SCM) is the process of planning, implementing, and controlling the business operations throughout, from product inception to delivery as efficiently as possible. Thus Supply chain activities cover everything from product development, sourcing, production, and logistics, as well as the information systems needed to coordinate these activities. SCM spans a movement and storage of raw materials, work in progress inventory, and finished goods from point of origin to point of consumption [3].

The development of a practical supply chain management (SCM) system can not only realize the high-efficiency of operations flows in business value chain but also can establish a new information management solution (IMS) for any industry. To achieve a SCM system, retailer needs to manage the supply chain effectively and apply IT to system such as communications technology, computer technology. In order to promote the SCM, retailer must establish a management system. System model must be established before the establishment of management system [5]. We have used unified modeling language to develop system for any industry in general and retail SME industry in particular. Any supply chain enterprise has the possibility to win in the competition only when it forms the strategic alliance with the upstream and downstream enterprise.

## 5. Analysis and Modelling of processes in retail system using UML

We try to automate the business process exist in any small and medium enterprises (SME) and covered the detailed supply chain transactions and Customer management information, this will speed up the supply chain process and manage customer information to forecast what their customer's trend will be in future particularly for manufacturing and retail systems.

### 5.1 Use Case Analysis

Use case diagrams address the business processes that the system will implement. Use cases describe the functional capabilities of the system and the external actors that interact with it [6]. Procure of goods is an important activity in any business system, normally this task is handled by retailer's purchasing department. The procurement process links members in the supply chain. Effective procurement contributes to the competitive advantage of a retailer. Typical procurement process includes the following stages:

(1) According to the user's need purchasing department set up purchasing plan and formulate expenditure which is delivered to financial department [6];

(2) Purchasing department transmit purchase documents to suppliers. Electronic data interchange (EDI), which involves the electronic transfer of purchase documents between the buyer and seller, can help shorten order cycle time. EDI transactions, particularly through the Internet, will increase over the next several years [7].

(3) Warehouse department receipt/inspection/in storage goods and delivery warehouse warrant to financial department.

(4) Financial department formulate account receivable according invoice.

Based on the process of procurement management, use case can be identified and procurement management model be drawn shown in figure 7.

### 5.2 Model for procurement management

This model represents the procurement process and emphasizes the flow of control among objects and models the function of a system. This models also simulates to activity diagram for procurement management

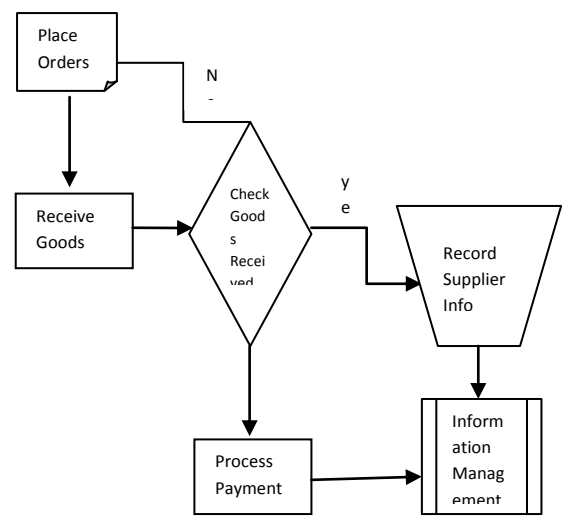

**Fig. 7 Procurement Process**

### 5.3 Class Diagram for Retailer System

The diagram below (figure 8) describes the retailer system and showcase how different types of object and data element involved in the system along with the relationships that exist among them. This class diagram is a conceptual diagram and is free from any structural implementations

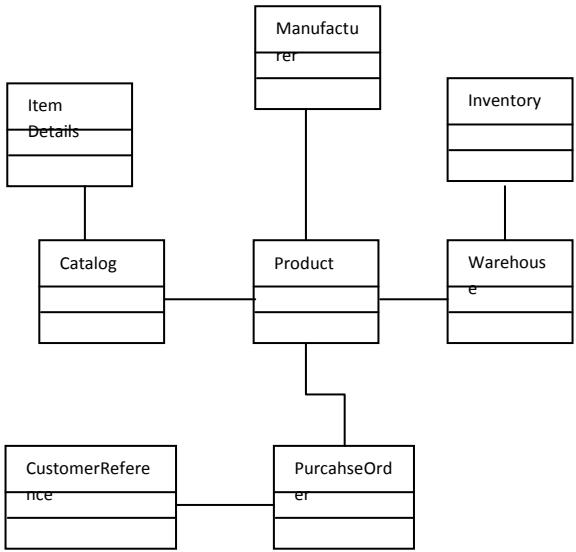

**Fig 8 Class Diagram for retailer systems**

## 5.4 Transaction Sequence of Goods Purchase

This is a dynamic model of a system, Systems dynamic behaviour can be described using UML dynamic modelling. UML dynamic models can be represented by sequence diagram, collaboration diagram, activity diagram and state chart diagram, these dynamic diagrams describes object behaviour & interactions between objects from different perspective. Below is the transaction sequence diagram for goods purchase (see figure 9)

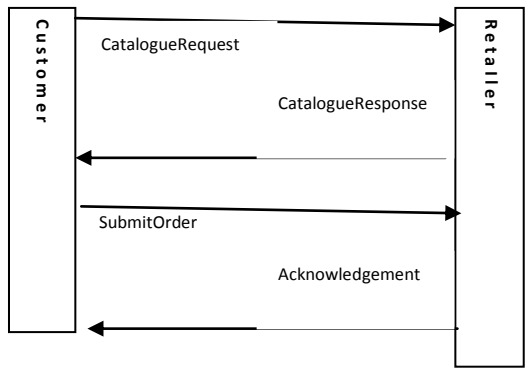

**Fig. 9 Transaction Sequence of Goods Purchase**

## 5.5 Transaction Sequence of Source Goods

This dynamic model describes the interaction between the retailer service and warehouse services. Retailer request to different warehouses for shipping goods, consequently the warehouse responds if it is in the capacity to fulfill the desired request (see figure 10)

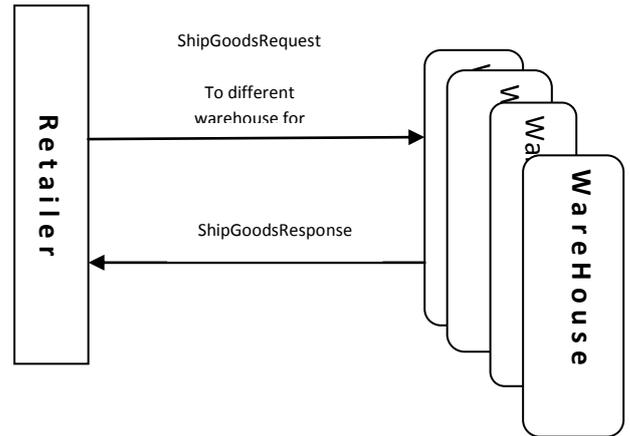

**Fig. 10 Transaction Sequence of Source Goods**

## 5.6 Transaction Sequence of Replenish Stocks

This dynamic model shows the interaction between warehouse and the manufacturer. If any of the warehouses is unable to fulfil the desired shipping request, it further interacts with the respective manufactures by submitting purchase order (POSubmit), manufacturer acknowledges in response to it. (See figure 11)

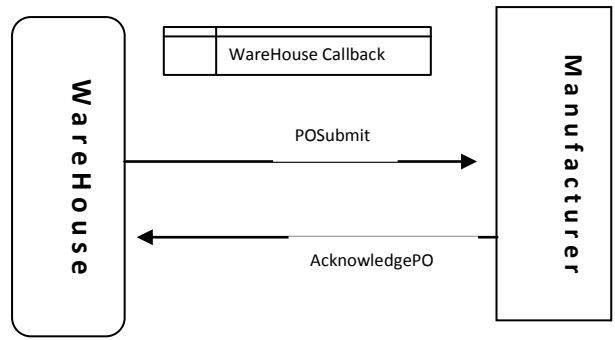

**Fig. 11 Transaction Sequence of Replenish Stocks**

## 6. Conclusions

In our work we have presented the operation and representation of three popular enterprise systems ERP, CRM and SCM with their respective objectives and architecture. It has been observed that in marketplace the maintenance of each enterprise systems independently is more costly and timely consuming than to develop an integrated system. Thus proposed diagrams enable business processes to be fully automated in between supplier, customer and the company.

## 7. References

[1] Fullerton M, Fernandez E.B ,Analysis Pattern for Customer relationship Management(CRM)

[2] Kumaran S, Bishop P, Chao T, Dhoolia P, Jain P, Jaluka R, Ludwig H., Moyer A, Nigam A, Using a Model Driven Transformational Approach and Service Oriented Architecture for Service Delivery Management, IBM Systems Journal Vol 46, No. 3 ,2007 pp513-519